\documentclass[conference]{IEEEtran}
\IEEEoverridecommandlockouts
\usepackage{cite}
\usepackage{amsmath,amssymb,amsfonts}
\usepackage{algorithmic}
\usepackage{graphicx}
\usepackage{textcomp}
\usepackage{xcolor}
\usepackage{hyperref}
\hypersetup{
colorlinks,
citecolor=blue,
linkcolor=blue,
urlcolor=blue,
}

\everydisplay\expandafter{\the\everydisplay
  \abovedisplayskip=3pt plus 1pt minus 1pt
  \belowdisplayskip=3pt plus 1pt minus 1pt
  \abovedisplayshortskip=0pt plus 1pt
  \belowdisplayshortskip=2pt plus 1pt}
\makeatletter
\def\section{\@startsection{section}{1}{\z@}{0.9ex plus 1ex minus 0.3ex}%
{0.5ex plus 0.5ex minus 0ex}{\normalfont\normalsize\centering\scshape}}
\def\subsection{\@startsection{subsection}{2}{\z@}{0.9ex plus 1ex minus 0.3ex}%
{0.5ex plus 0.3ex minus 0ex}{\normalfont\normalsize\itshape}}
\makeatother
\def\BibTeX{{\rm B\kern-.05em{\sc i\kern-.025em b}\kern-.08em
    T\kern-.1667em\lower.7ex\hbox{E}\kern-.125emX}}
\begin{document}

\title{Probabilistic Power Estimation and MPC-based Scheduling of Construction Electric Vehicles Supported by Mobile Charging Stations\\
\thanks{* S.~Mital and A.~Ghosh contributed equally. \\
This work was supported by the California Energy Commission under the award number EPC-24-031.\\
Email:\{smital,avghosh,d2rojas,sdey,yyshi\}@ucsd.edu}
}

\author{\IEEEauthorblockN{Shubhan Mital$^*$, Avik Ghosh$^*$, Daniela Rojas, Sujit Dey, and Yuanyuan Shi}
\IEEEauthorblockA{\textit{Department of Electrical and Computer Engineering} \\
\textit{University of California, San Diego, La Jolla, CA 92093, USA}\\ }}


\maketitle

\begin{abstract}
The adoption of construction electric vehicles (CEVs) 
is held back by sparse on-site charging infrastructure, limited mobility, and insufficient knowledge of their power consumption. Mobile charging stations (MCSs) towed between grid connection points and construction sites address the first barrier, yet scheduling them optimally requires an accurate 
estimate of the CEV power consumption. In this paper, we address these gaps, first, by estimating the power of each subactivity of a representative CEV (a compact electric excavator) based on real-world construction demonstration data. We use Bayesian linear regression to obtain the subactivity power distributions, accounting for their operational uncertainty. Second, we formulate a mixed-integer program that jointly schedules CEV work and charging together with MCS routing, charging, and discharging, optimizing electricity costs, carbon emissions, work completion, and MCS travel. We embed the optimizer in a shrinking-horizon model predictive control (MPC) loop to perform real-time control on the CEV-MCS system. Two MPC variants, one a certainty-equivalent MPC (CE-MPC) controller, and the other a scenario-based MPC (SB-MPC) controller, are compared in closed-loop against an open-loop day-ahead schedule. 
Realistic simulations show that SB-MPC performs the best, lowering the operating costs by $60.2\%$ relative to the open-loop baseline, and by $23.3\%$ relative to CE-MPC, demonstrating the efficacy of incorporating predictive control with feedback. 
The code is available at \url{https://github.com/Shubhanflash22/CEV-MCS-MPC-Joint-Scheduling}.\\
\end{abstract}


\section{Introduction}\label{sec:intro}
\subsection{Motivation}\label{subsec:motivation}
Diesel-powered construction equipments are being increasingly recognized as significant contributors to lifecycle emissions and local
air pollution~\cite{desouza2020real}. Public buyers are turning this concern into procurement rules: Oslo requires all municipal construction sites to be emission-free from 2025~\cite{bigbuyers2022zemcons}, while California subsidizes zero-emission off-road equipment~\cite{core}. However, the widespread adoption of construction EVs (CEVs) is hindered by the following problems: (i) setting up charging infrastructure at a construction site is expensive and infeasible as a construction site is a temporary load~\cite{bigbuyers2022zemcons}; (ii) CEVs are typically immobile and cannot travel to stationary charging infrastructure, as on-road travel of CEVs is often hindered by construction site conditions, safety issues, and the equipment itself~\cite{lin2020development}; and even if travel is possible, (iii) synchronized CEV fleet charging at EV charging stations can overload primary substations~\cite{jansson2025charging}. Construction operators can therefore deploy mobile charging stations (MCSs), i.e., a large battery towed between grid-connection points and construction sites, to decouple CEV charging from fixed charging infrastructure~\cite{shin2026assessment}. This decoupling can help exploit temporal flexibility in grid power draw by shifting MCS grid charging to low-cost and low-carbon periods. 

In a recent work~\cite{cev_mcs_extended} by the authors, we showed that exploiting this flexibility well requires co-optimizing \emph{when and which subactivities} each CEV performs, \emph{when and how much} each CEV charges, with \emph{where, when, and how much} its MCS charges/discharges. However, the framework in~\cite{cev_mcs_extended} is deterministic as the day's CEV-MCS schedule is computed once, before the construction work shift, and implemented in open-loop. The planner creates the optimized schedule by fixing the power of each subactivity to a point estimate. An open-loop plan built on a point estimate of subactivity powers can leave a CEV without charge before its work is complete and may fail to adjust its MCS recharging schedule in response to the deviation of real-time CEV power usage from its point estimate. This paper extends~\cite{cev_mcs_extended} to make the schedule aware of, and responsive to, this uncertainty in CEV subactivity power, both from a planning and closed-loop control point of view.

\subsection{Literature Review}\label{subsec:lit_review}
Studies on power consumption models of CEVs mostly focus on powertrain design~\cite{huang2024prospects}, optimal sizing and power capacity ratings of CEV batteries~\cite{unnoor2023battery}, or more recently on MCS infrastructure design~\cite{shin2026assessment}. These studies, however, do not analyze CEV power demand at a subactivity (e.g., digging, loading, etc.) level granularity, which is important for optimal work-charging scheduling. Our prior work~\cite{cev_mcs_extended} recovered deterministic point estimates of various subactivity powers of CEVs based on a real-world construction demonstration. 
To the best of our knowledge, no probabilistic, subactivity-resolved power consumption model exists for CEVs.

As studies on optimal scheduling of CEV-MCS systems are limited, we next analyze the value of incorporating uncertainty in scheduling similar electric freight vehicles, trucks, and buses. Robust routing of electric freight vehicles in~\cite{pelletier2019evrp} guards against stochastic consumption parameters by trading off risk with cost. Ignoring consumption uncertainty in electric bus scheduling leaves up to $48\%$ of trips delayed or missed in~\cite{avishan2023fleet}. While the previous plans are open-loop, re-planning (i.e., closed-loop control with feedback) from measured state-of-charge (SOC) has been adopted for electric trucks that update charging plans en route~\cite{bai2025distributed} and bus depots that track a day-ahead plan with a receding-horizon planner under time-of-use and demand charges~\cite{whitaker2025scheduling}. These studies demonstrated that closed-loop control added performance and robustness benefits over open-loop approaches. However, existing CEV work-charging scheduling approaches~\cite{huang2024energy} do not consider uncertainty and are open-loop.

MCSs have been dispatched for EV charging demand requests incorporating uncertainty by iteratively learning from historical data in~\cite{he2026coordinated}, and by reinforcement learning in~\cite{do2026sustainable}. 
In all of these studies, the mobile asset serves an exogenous demand, whereas for optimal CEV-MCS operation, the demand is the work schedule itself, decided jointly with the charging. Our previous work~\cite{cev_mcs_extended} jointly optimized CEV work scheduling and MCS charging, but assumed deterministic CEV power consumption and solved the problem open-loop. To the best of our knowledge, no prior work couples a probabilistic CEV power consumption model with uncertainty-aware, closed-loop optimization of CEV work and MCS charging.

\subsection{Contributions}\label{subsec:contributions}
The contributions of this work are as follows:
\begin{enumerate}
\item We develop a Bayesian linear regression model for CEV power consumption that yields a \emph{probability distribution of each subactivity's power draw}. Unlike prior models that give deterministic power estimates, our model explicitly quantifies prediction uncertainty, enabling work and charging optimization that hedges against power consumption variations, a capability no prior work provides.

\item We propose an \emph{uncertainty-aware, closed-loop framework} for joint CEV work and MCS charging optimization, realized through two model predictive control (MPC) variants: a certainty-equivalent MPC and a scenario-based MPC. Unlike the open-loop, deterministic approach in~\cite{cev_mcs_extended}, both proposed MPC methods handle power uncertainty in operation and through feedback.

\item We demonstrate that the proposed probabilistic power estimation model and closed-loop MPC framework can reduce the total system operating cost by \emph{up to $60.2\%$} compared to the prior open-loop baseline, showing the value of combining probabilistic power estimates with uncertainty-aware predictive control.
\end{enumerate}

The rest of the paper is organized as follows. Section~\ref{sec:power} presents the Bayesian regression based power estimation, Section~\ref{sec:formulation} presents the MPC-based joint scheduling strategies of the CEV-MCS system, Section~\ref{sec:results} presents the case study setup and results, and Section~\ref{sec:conclusions} concludes the study.

\section{Bayesian Linear Regression for Subactivity Power Estimation}\label{sec:power}





Based on a real-world construction demonstration involving soil work at UCSD with a compact electric  excavator, we determine four {separately identifiable} subactivity classes. The four CEV subactivities are defined by the set $\mathcal{A}=\{\text{digging, loading$+$swinging, traveling, idling}\}$. In this section, we first motivate the Bayesian linear regression framework for CEV subactivity power estimation in Section~\ref{subsec:energybalance}, then present its formulation in Section~\ref{subsec:bayes}, and finally validate it in Section~\ref{subsec:power_validation}. 
%


\subsection{Energy-balance Model}\label{subsec:energybalance}
Scheduling CEV work and MCS charging requires predicting the CEV's energy consumption, which varies across executions of the same subactivity and must be inferred from limited field data. A deterministic power value per subactivity can therefore under- or overestimate consumption, leading to battery depletion or unnecessary charging. We thus estimate a probability distribution over each subactivity's power, using the real-world field data from~\cite{cev_mcs_extended}: videos of the CEV labeled with the subactivity performed at each second and time-synchronized with the SOC reported by its telematics. These data are partitioned into \emph{quantization-aware} windows indexed by $i \in \{1,\dots,d\}$, relating the time spent on each subactivity to the energy consumed as
\begin{equation}
b_i = \sum_{a \in \mathcal{A}} p_a D_{i,a} + \varepsilon_i,
\label{eq:balance_window}
\end{equation}
\noindent where $b_i$ (kWh) is the energy consumed by the CEV, $p_a \ge 0$ (kW) is the mean power of subactivity $a\in\mathcal{A}$, $D_{i,a}$ (h) is the time spent on subactivity $a$, and $\varepsilon_i$ (kWh) is the residual. Stacking all $d$ windows gives $\mathbf{b} = \mathbf{D}\mathbf{p} + \boldsymbol{\varepsilon}$. Unlike the non-negative least squares (NNLS) fit of~\cite{cev_mcs_extended}, which returns a point power estimate per subactivity, we treat $\mathbf{p}$ as a random vector and infer its posterior distribution jointly with the residual standard deviation.

\subsection{Bayesian Formulation}\label{subsec:bayes}
We infer $\mathbf{p}$ and the standard deviation $\varsigma$ of the residual jointly from the observed window energies $\mathbf{b}$ by Bayesian linear regression~\cite{basso2021evrp}, i.e., by Bayes' rule,

\small{
\begin{equation}
\underbrace{\pi(\mathbf{p}, \varsigma \mid \mathbf{b})}_{\text{posterior}} \;\propto\; \underbrace{\mathcal{L}(\mathbf{b} \mid \mathbf{p}, \varsigma)}_{\text{likelihood}}\;\underbrace{\pi(\mathbf{p})\,\pi(\varsigma)}_{\text{prior}}.
\label{eq:bayes}
\end{equation}}
\normalsize

\noindent In~\eqref{eq:bayes}, the prior encodes the initial guess about $\mathbf{p}$ and $\varsigma$, the likelihood is the probability density of $\mathbf{b}$ for a given $\mathbf{p}$ and $\varsigma$, and the posterior encodes the updated joint probability distribution of $\mathbf{p}$ and $\varsigma$ after learning from the observed data. The priors of the individual $p_a$ and of $\varsigma$ are assumed to be independent. Note that $\mathcal{N}(\mu, s^2)$ denotes a normal distribution with mean $\mu$ and variance $s^2$, $\mathcal{N}(x; \mu, s^2)$ denotes its probability density evaluated at $x$, and $\mathcal{N}^{+}(\mu, s^2)$ denotes the same normal distribution truncated to $[0,\infty)$.

\subsubsection{Likelihood}\label{subsubsec:likelihood} 

We model the residuals as independent zero-mean normal random variables with a common standard deviation, i.e., $\varepsilon_i \sim \mathcal{N}(0, \varsigma^2)$, 
so that~\eqref{eq:balance_window} gives

\small{
\begin{equation}
\mathcal{L}(\mathbf{b} \mid \mathbf{p}, \varsigma) = \prod_{i=1}^{d} \mathcal{N}\Big(b_i;\ \textstyle\sum_{a} p_a D_{i,a},\ \varsigma^2\Big).
\label{eq:likelihood}
\end{equation}}
\normalsize

\subsubsection{Prior}\label{subsubsec:prior}

Since the power drawn by a subactivity cannot be negative, each $p_a$ has a truncated normal prior, i.e., $p_a \sim \mathcal{N}^{+}\big(\bar{p}^{0}_a, (\sigma^{0}_a)^2\big)$. We set the prior mean $\bar{p}^{0}_a$ to the NNLS subactivity power estimate and the prior standard deviation $\sigma^{0}_a$ to the standard deviation of the NNLS estimate across repeated 5-fold cross-validation fits. 
The residual standard deviation also has a truncated normal prior, i.e., $\varsigma \sim \mathcal{N}^{+}(0, \sigma_b^2)$, where the prior standard deviation $\sigma_b$ is set to the standard deviation of the observed CEV energy consumption $\mathbf{b}$.




Because of the truncated priors, the posterior~\eqref{eq:bayes} has no closed form. Thus, we represent it by samples generated by Markov chain Monte Carlo using the No-U-Turn sampler~\cite{hoffman2014nuts}.

\subsection{Subactivity Power Validation of the Bayesian Formulation}\label{subsec:power_validation}
We validate the Bayesian linear regression fit with Pareto-smoothed importance-sampling leave-one-out cross-validation (PSIS-LOO)~\cite{vehtari2017psis}. PSIS-LOO is the Bayesian counterpart of the NNLS leave-one-out (NNLS-LOO) cross-validation from the point estimates of subactivity power. For each window, PSIS-LOO estimates the energy predicted through the posterior fitted without that window. For PSIS-LOO, the mean absolute error (MAE) and mean absolute percentage error (MAPE) in energy are $0.119$~kWh and $16.9\%$, respectively, slightly lower than the $0.13$~kWh and $18\%$ obtained by NNLS with LOO cross-validation. The Bayesian model thus matches the accuracy of NNLS while additionally quantifying uncertainty. The implementation and the resulting posterior distributions of the subactivity powers are publicly available in our GitHub repository~\cite{cevmcs_mpc_code}. 

\section{MPC-based joint scheduling of the CEV-MCS system}\label{sec:formulation}
This section formulates the joint optimization framework for CEV work-charging and MCS charging-discharging using the posterior distributions of the subactivity powers from Section~\ref{sec:power}. We propose two MPC strategies for the joint scheduling: a certainty-equivalent MPC (CE-MPC), and a scenario-based MPC (SB-MPC) approach. Both MPC strategies are implemented in a shrinking-horizon fashion. 

Both MPC strategies consist of an open-loop planning stage (Sections~\ref{subsec:objective} and~\ref{subsec:constraints}) followed by a closed-loop implementation stage (Section~\ref{subsec:strategies}) with feedback. In the planning stage, using the subactivity powers from Section~\ref{sec:power} as an exogenous input, we solve a finite-horizon optimal control problem from the current time-step and state to predict the schedule of the CEV-MCS system over the remaining horizon. In the implementation stage, we apply only the first control input of this schedule at the current time-step, and the system state evolves to the next time-step under the applied control input and the realized disturbances. The process then repeats from the new state and (the next) time-step, with the remaining  horizon shortened by one time-step. In the planning stage optimization, the SB-MPC samples $S=5$ scenarios from the posterior subactivity power distributions, whereas the CE-MPC uses a single scenario ($S=1$) equal to the posterior mean of each subactivity power.

We consider a work-charging network of construction nodes (denoted by the set $\mathcal{N}^c$) and grid connection nodes (denoted by the set $\mathcal{N}^g$). Let $\mathcal{E}$ be the set of CEVs, each CEV being assigned to a fixed construction node, and $\mathcal{M}$ be the set of MCSs, with each MCS shuttling between grid connection nodes and construction nodes. The set of all nodes is denoted by $\mathcal{N} = \mathcal{N}^g \cup \mathcal{N}^c$. Time is discretized into intervals (i.e., time-steps) denoted by $t \in \mathcal{T} = \{1,\dots,t_{\mathrm{end}}\}$ of length $\Delta T$, with 
$\mathcal{T}^{\mathrm{op}} \subseteq \mathcal{T}$ denoting on-peak ($4-9$ pm) hours. At the current time-step $t_0$, the planning stage covers the remaining horizon consisting of the  time-steps denoted by the set $\mathcal{K}(t_0)=\{t_0,\dots,t_{\mathrm{end}}\}$.

\subsection{Objective Function for the Planning Stage}\label{subsec:objective}
The optimization for the planning stage minimizes a weighted cost over the subactivity power scenarios. The objective function at the current time-step $t_0$ is formulated as 

\small{
\begin{align}
J(t_0) ={}& \sum_{s=1}^{S} w_s \Bigg[ \Bigg\{\sum_{m \in \mathcal{M}}\sum_{t \in \mathcal{K}(t_0)} \big(\lambda^{\mathrm{elec}}_t + \lambda^{\mathrm{CO_2}}_t \lambda^{\mathrm{em}}\big) P^{\mathrm{ch,tot}}_{m,t,s}\Delta T \nonumber\\
&\qquad \qquad + \lambda^{\mathrm{NC}} P^{\mathrm{NC}}_s(t_0) + \lambda^{\mathrm{OP}} P^{\mathrm{OP}}_s(t_0) \nonumber\\
&\qquad \qquad + \rho^{\mathrm{miss}} \sum_{i \in \mathcal{N}^c}\sum_{a \in \mathcal{A}} s^{\mathrm{miss}}_{i,a,s} \nonumber\\
&\qquad \qquad + \rho^{\mathrm{travel}}\Delta T \sum_{m \in \mathcal{M}}\sum_{i \in \mathcal{N}}\sum_{j \in \mathcal{N}}\sum_{t \in \mathcal{K}(t_0)} y_{m,i,j,t,s} \nonumber\\
&\qquad \qquad + \rho^{\mathrm{short}} \sum_{i \in \mathcal{N}^c}\sum_{e \in \mathcal{E}} s^{\mathrm{short}}_{i,e,s}\Bigg\} \nonumber\\
&\qquad \quad + \epsilon^{\mathrm{tb}} \sum_{i \in \mathcal{N}^c}\sum_{e \in \mathcal{E}}\sum_{t \in \mathcal{K}(t_0)} (t-t_0+1)\,\mu_{i,e,t,s} \Bigg].
\label{eq:objective}
\end{align}}
\normalsize

\noindent The outermost sum in~\eqref{eq:objective} averages the operating cost (terms within braces), 
with each scenario $s$ having weight $w_{s}$. We give equal weights to each scenario, expressed by $w_{1}=w_{2}=\cdots=w_{S}$ and $\sum_{s=1}^{S} w_s = 1$.

The first two terms in~\eqref{eq:objective} are the grid electricity and carbon emissions costs of charging the MCSs; respectively. $P^{\mathrm{ch,tot}}_{m,t,s}$ (kW) is the grid charging power of MCS $m$ at time-step $t$ in scenario $s$, $\lambda^{\mathrm{elec}}_t$ (\$/kWh) and $\lambda^{\mathrm{CO_2}}_t$ (kgCO$_2$/kWh) are the grid electricity rate and carbon intensity at time-step $t$; respectively, and $\lambda^{\mathrm{em}}$ (\$/kgCO$_2$) converts carbon emissions to dollar terms. 

The third and fourth terms in~\eqref{eq:objective} are the non-coincident (NC) and on-peak (OP) demand charges, with rates $\lambda^{\mathrm{NC}}$ (\$/kW) and $\lambda^{\mathrm{OP}}$ (\$/kW); respectively. The NC demand charge is a penalty on the NC demand peak, i.e., the maximum power drawn from the grid over all time intervals; while the OP demand charge is a penalty on the OP demand peak, i.e., the maximum power drawn from the grid over on-peak hours. 
The NC demand peak in scenario $s$ is defined as $P^{\mathrm{NC}}_s(t_0)\!\!=\!\!\max\{P^{\mathrm{NC}}(t_0\!-\!1),\max_{t\in\mathcal{K}(t_0)}\sum_m P^{\mathrm{ch,tot}}_{m,t,s}\}$, where $P^{\mathrm{NC}}(t_0\!-\!1)$ (kW) is the closed-loop NC peak realized before $t_0$. The OP peak is similarly defined as $P^{\mathrm{OP}}_s(t_0)=\max\{P^{\mathrm{OP}}(t_0-1),\max_{t\in\mathcal{K}(t_0)\cap\mathcal{T}^{\mathrm{op}}}\sum_m P^{\mathrm{ch,tot}}_{m,t,s}\}$, where $P^{\mathrm{OP}}(t_0\!-\!1)$ (kW) is the closed-loop OP peak realized before $t_0$. 

The fifth term in~\eqref{eq:objective} penalizes the unfinished work $s^{\mathrm{miss}}_{i,a,s}$ (h) of subactivity $a\in\mathcal{A}$ at construction node $i$ in scenario $s$. The penalty $\rho^{\mathrm{miss}}$ (\$/h) is set two orders of magnitude higher than the other cost coefficients so that timely work completion gets the highest priority. The sixth term in~\eqref{eq:objective} is the labor cost of towing the MCSs, where the binary variable $y_{m,i,j,t,s}$ equals 1 if and only if MCS $m$ is in transit from node $i$ to node $j$ at time-step $t$ for scenario $s$, and $\rho^{\mathrm{travel}}$ (\$/h) is the labor cost. The seventh term in~\eqref{eq:objective} penalizes the terminal energy shortfall $s^{\mathrm{short}}_{i,e,s}$ (kWh) by which the battery state of energy (SOE) of CEV $e$ at node $i$ at $t_{\mathrm{end}}$ falls below its initial value, since an undercharged CEV disrupts the next day's work. Its penalty $\rho^{\mathrm{short}}=\rho^{\mathrm{miss}}/\bar{P}^{\mathrm{work}}$ (\$/kWh), where $\bar{P}^{\mathrm{work}}$ (kW) is the average work power, charges the shortfall as equivalent hours of unfinished work.

The last term in~\eqref{eq:objective} is a tie-breaker that prefers earlier CEV charging, where the binary variable $\mu_{i,e,t,s}$ equals 1 if and only if CEV $e$ at node $i$ is connected to an MCS at time-step $t$ in scenario $s$ for charging. Its weight $(t-t_0+1)$ increases with time, so later charging costs slightly more, and the weight $\epsilon^{\mathrm{tb}} = 10^{-3}$ is set small enough, so that it does not alter the schedule affecting the operating costs in~\eqref{eq:objective}.

\subsection{Constraints for the Planning Stage}\label{subsec:constraints}
The planning-stage optimization has four types of constraints, which are the same as those of~\cite{cev_mcs_extended}. Due to the page limit, we only summarize the constraints here and refer the reader to~\cite[Sec.~III-B]{cev_mcs_extended} for the detailed formulation. What differs in the proposed framework is: the constraints are imposed for every power scenario $s$, so the plan accounts for uncertainty in subactivity power, and the problem is re-solved at each MPC step from the measured SOE, closing the loop that the open-loop, deterministic controller used in \cite{cev_mcs_extended}.

\emph{1) Charging and discharging power constraints of the MCSs and CEVs} 
keep the charging and discharging powers of the MCSs and the charging power of the CEVs within their rated limits, ensure that an MCS charges only at grid nodes and discharges only at construction nodes, and require that the energy an MCS discharges is received by its plugged-in CEVs.

\emph{2) Energy consumption and SOE constraints of the MCSs and CEVs} 
ensure that every CEV works only at its assigned construction node during working hours, that CEV working and charging are mutually exclusive, and that the SOE of every MCS and CEV evolves with the energy charged, discharged and consumed, accounting for efficiencies, and stays within its operating limits. They also return each MCS and CEV to their initial SOE at the end of the horizon. Note that due to different powers of the subactivities in closed-loop than their open-loop planning stage, the final SOE of the CEV might be different from its initial (i.e., $s^{\mathrm{short}}_{i,e,s} \neq 0$ in closed-loop despite $s^{\mathrm{short}}_{i,e,s} = 0$ in open-loop). 

\emph{3) Spatial routing and connection status constraints of the MCSs and CEVs} 
keep each MCS either parked at one node or in transit on one route, track its arrivals and departures given the fixed travel times, and let a CEV connect only to an MCS present at its node.

\emph{4) Operational adjustment constraints for CEV work scheduling} 
ensure that each CEV does at most one subactivity per time-step, and require the subactivity durations to be completed up to the maximum allowed slack $s^{\mathrm{miss}}_{i,a,s}\ge 0$. The constraints also enforce CEV precedence (i.e., digging before loading+swinging), rest (i.e., idle the CEV after a certain amount of continuous work), and travel-pacing (i.e., reposition the CEV upon finishing a certain amount of work) rules.

Let $\mathcal{D}_{t_{0},s}$ be the set of decision variables of our optimization for the current time-step $t_{0}$ 
for scenario $s$. The control that is applied in closed-loop at $t_{0}$ is constrained to be identical in every scenario (i.e., $\mathcal{D}_{t_0,s}$ is the same $\forall s \in \{1,2,\dots,S\}$). The decisions for later time-steps (i.e., $\mathcal{D}_{t_0+1,s},\dots, \mathcal{D}_{t_{\mathrm{end}},s}$) stay free in each scenario $s$. For CE-MPC, which has a single scenario, this coupling is trivial.

\subsection{Closed-loop Implementation Stage}
\label{subsec:strategies}

\emph{Open-loop baseline.} The open-loop baseline~\cite{cev_mcs_extended} solves the planning problem in Section~\ref{subsec:objective}-\ref{subsec:constraints} at $t_0 = 1$ with a single scenario in which each subactivity power equals its posterior mean. Then it executes that plan for the entire horizon.

\emph{MPC-based strategies.} For both MPC schemes, we solve the planning problem in Section~\ref{subsec:objective}-\ref{subsec:constraints} at every time-step $t_0 \in \{1,\dots,t_{\mathrm{end}}\}$ from the current state and apply only the first control input at $t_0$. The CEV-MCS system then realizes the closed-loop subactivity powers, which in general differ from the planned ones, and updates the CEV and MCS states, and the outstanding work requirements. 
The time-step then advances, $t_0\! \leftarrow\! t_0\!+\!1$, the remaining horizon shrinks by one time-step, and the MPC problem is re-solved based on the updated CEV, MCS, and remaining work states. 

\emph{Synthetic database for closed-loop subactivity powers.} A synthetic database of ground-truth subactivity powers is created from the field data of CEV power consumption. For each window $i$ of Section~\ref{subsec:energybalance}, we find the subactivity powers closest to their posterior means $\bar{p}_a$ that reproduce the measured energy drop exactly. Concretely, we minimize $\sum_{a\in\mathcal{A}}(p_a-\bar{p}_a)^2$ subject to $p_a\ge 0$ and~\eqref{eq:balance_window} with $\varepsilon_i=0$, and record the resulting $p_a$ of every subactivity present in the window. 

\section{Results and Discussion}\label{sec:results}

\textbf{Case study setup.} Mirroring the UC San Diego field demonstration, we simulate the CEV-MCS system with one CEV and one MCS that shuttles between the grid node and a construction site. The CEV and MCS specifications, scheduling horizon, cost parameters, electricity and carbon tariffs, and work order are the same as those of~\cite{cev_mcs_extended} for a fair comparison. 
In closed-loop, the CEV-MCS system samples the power of each subactivity from the synthetic database of Section~\ref{subsec:strategies}. The controllers do not see the synthetic database and plan only with the Bayesian posterior of the subactivity power estimate derived in Section~\ref{sec:power}. We run $100$ seeds, each a different random draw of the closed-loop subactivity powers, and all strategies face the same $100$ draws. Each optimization is solved with Gurobi~\cite{gurobi} under a $600$~s time limit. 

\begin{table}[t]
\centering
\caption{Total cost and unfinished work penalty comparison of the open-loop, CE-MPC and SB-MPC controllers over 100 seeds.}
\label{tab:case}
\footnotesize
\begin{tabular}{lrrr}
\hline
 & Open-loop & CE-MPC & SB-MPC \\
\hline
Mean operating cost (\$) & 280 & 145 & \textbf{111} \\
Unfinished work penalty (\$) & \textbf{4} & 75 & 42 \\
Terminal shortfall penalty (\$) & 207 & \textbf{0} & \textbf{0} \\
No. of seeds with unfinished work & 8 & 17 & \textbf{11} \\
No. of seeds with terminal shortfall & 66 & \textbf{0} & \textbf{0} \\
\hline
\end{tabular}
\end{table}


\textbf{Results based on the synthetic database.} Table~\ref{tab:case} shows that SB-MPC lowers the mean operating cost by $60.2\%$ relative to the open-loop baseline, and by $23.3\%$ relative to CE-MPC, while CE-MPC lowers it by $48.1\%$ relative to the open-loop baseline. The biggest gap across the open-loop and MPC strategies comes from the terminal shortfall of CEV SOE, which both MPC strategies eliminate by incorporating feedback. The gap between the MPC strategies arises from differences in unfinished work, whose penalty is lower under SB-MPC relative to CE-MPC. This shows the benefit of planning against scenarios (in SB-MPC) rather than simply reacting to it (in CE-MPC) in operating cost savings.


\textbf{Sensitivity analysis.} To test the robustness of our controllers, we repeat the case study with the synthetic database of Section~\ref{subsec:strategies} replaced by three modes, in which the closed-loop power of each subactivity is $\max(\mu+\sigma z,0)$, where $\mu$ and $\sigma$ are its posterior mean and standard deviation and $z$ is the deviation from the mean in standard deviations (Table~\ref{tab:sens}). The \emph{near mean} mode keeps the closed-loop subactivity power close to the posterior means, while the \emph{high} and \emph{low} modes place it closer to the tails of the posterior distribution. Each mode is run for 10 seeds with all three strategies.

\begin{table}[t]
\vspace{-0.5em}
\centering
\caption{Mean operating cost (\$) by power mode (n=10 seeds).}
\label{tab:sens}
\footnotesize
\begin{tabular}{llrrr}
\hline
Mode & $z$ & Open-loop & CE-MPC & SB-MPC \\
\hline
Low & $-2-0.5|\xi|$ & 67 & 64 & \textbf{63} \\
Near mean & $0.5\,\xi$ & 81 & \textbf{69} & \textbf{69} \\
High & $2+0.5|\xi|$ & 834 & 475 & \textbf{375} \\
\hline
\multicolumn{5}{p{0.9\columnwidth}}{\scriptsize Here $\xi\sim\mathcal{N}(0,1)$ is a standard normal random draw.}
\end{tabular}
\vspace{-0.5em}
\end{table}

Table~\ref{tab:sens} shows that in the low and near mean modes, both MPC strategies cost less than the open-loop baseline and perform essentially on par. In the low mode SB-MPC is marginally cheaper than CE-MPC, and in the near mean mode the two are equal. In the high mode the benefit of planning against multiple scenarios becomes clear. CE-MPC reduces the cost relative to the open-loop baseline by $43.0\%$, while SB-MPC reduces it by $55.0\%$. 

\section{Conclusions}\label{sec:conclusions}
This paper develops a probabilistic model of CEV subactivity power to jointly schedule CEV work-charging and MCS charging-discharging in real-time. A Bayesian linear regression based on labeled field videos and SOC telematics gives a posterior distribution for the power of each subactivity, accurate to within $17\%$ mean absolute percentage error. We feed the posterior subactivity power distributions into two shrinking-horizon MPC variants: a scenario-based (SB-), and a certainty-equivalent (CE-) MPC. Realistic simulations on a stochastic CEV-MCS system show that both the MPC variants outperform an open-loop baseline by $48.1\%-60.2\%$ by incorporating feedback. Between the MPC strategies, the SB-MPC offers additional cost savings over the CE-MPC by planning against multiple scenarios. The code and data for all three approaches are released publicly in~\cite{cevmcs_mpc_code}.


\bibliographystyle{IEEEtran}
\bibliography{refs}

\end{document}